\documentclass[prl,a4paper,twocolumn,preprintnumbers,nofootinbib]{revtex4}
\usepackage[a4paper, hdivide={1.91cm,,1.165cm}, vdivide={1.83cm,,3.8cm}]{geometry}

\usepackage{amsmath,amssymb}
\usepackage{graphicx,multirow}
\usepackage{color}
\usepackage{slashed}
\usepackage{units}
\usepackage[hyperfootnotes=false]{hyperref}

\def \vec#1{{\boldsymbol{#1}}}
\newcommand{\hc}{\ensuremath{\text{h.c.}}}

\allowdisplaybreaks

\begin{document}

\preprint{ULB-TH/17-22}

\title{Proton decay into charged leptons}

\author{Thomas Hambye}
\email{thambye@ulb.ac.be}
\affiliation{Service de Physique Th\'eorique, Universit\'e Libre de Bruxelles, Boulevard du Triomphe, CP225, 1050 Brussels, Belgium}

\author{Julian Heeck}
\email{Julian.Heeck@ulb.ac.be}
\affiliation{Service de Physique Th\'eorique, Universit\'e Libre de Bruxelles, Boulevard du Triomphe, CP225, 1050 Brussels, Belgium}

\hypersetup{pdftitle={Proton decay into charged leptons},   pdfauthor={Thomas Hambye, Julian Heeck}}


\begin{abstract}
We discuss proton and neutron decays involving three leptons in the final state. Some of these modes could constitute the dominant decay channel because they conserve lepton-flavor symmetries that are broken in all usually considered channels. This includes the particularly interesting and rarely discussed $p\to e^+ e^+ \mu^-$ and $p\to \mu^+ \mu^+ e^-$ modes.  As the relevant effective operators arise at dimension 9 or 10, observation of a three-lepton mode would probe energy scales of order 100\,TeV. This allows to connect proton decay to other probes such as rare meson decays or collider physics. UV completions of this scenario involving leptoquarks unavoidably violate lepton flavor universality and could provide an explanation to the recent $b\to s\mu\mu$ anomalies observed in $B$ meson decays.
\end{abstract}

\maketitle


\section{Introduction}

The search for proton decay (PD) is one of the most important experimental endeavors in particle physics.
The proton is expected to be unstable  because baryon number is only an accidental symmetry in the Standard Model (SM), violated in many SM extensions~\cite{Nath:2006ut}. 
From the low-energy SM perspective, PD can be induced already at the level of dimension $d=6$ operators such as $uud\ell/\Lambda^2$~\cite{Weinberg:1979sa,Wilczek:1979hc}. This leads to two-body decays like $p\to \ell^+ \pi^0$ with rate $\Gamma \propto m_p^5/\Lambda^4$. As of today, the experimental sensitivity to such decays is of order $10^{34}$~years~\cite{Miura:2016krn}, so PD searches are currently probing an effective UV scale~$\Lambda$ of order $\unit[10^{15}]{GeV}$, i.e.~nothing but the GUT scale.

More generally PD could also test physics at a lower scale if the transition proceeds through an operator of dimension higher than 6. Without fine-tuning, this requires a symmetry that eliminates the lower-dimensional operators. 
On the basis of baryon number $B$ and lepton number $L$ symmetries, the corresponding list of dominant operators has been determined by Weinberg~\cite{Weinberg:1980bf}. Dominant here means the lowest-dimensional operators that conserve a given symmetry of the form $B+a L$ with $a\in \mathbb{Q}$.

In this letter we show that by adopting lepton \emph{flavor} symmetries instead of only $B$ and $L$, the list of operators which emerges is totally different, leading to different dominant decay modes. Many of these channels have not been discussed in the literature before and/or not been searched for experimentally, but could be probed very efficiently, e.g.~by Super-Kamiokande (SK). Due to the higher operator dimension, these channels are sensitive to scales down to $\unit[100]{TeV}$, which could have interesting associated signatures in other observables. 

By lepton flavor symmetries we mean combinations of the three \emph{individual} lepton flavor numbers $L_{e,\mu,\tau}$. These are conserved quantum numbers in the SM and have been observed to be broken only very weakly in the neutral lepton sector through neutrino oscillations.
As a result, flavor is still an excellent approximate symmetry in the charged lepton sector, up to unobservable neutrino-mass suppressed effects~\cite{Heeck:2016xwg}. 
PD operators up to dimension~8 involve only a single lepton, say of flavor $\alpha$, and thus simply violate $\Delta B=\pm \Delta L_\alpha$ and conserve $B\mp L$, $L_\beta$, and $L_\gamma$, with $\alpha$, $\beta$, $\gamma$ all different flavors.\footnote{Note that other kinds of horizontal symmetries were already qualitatively discussed in Refs.~\cite{Zee:1981vr,Sarkar:1983tm} for dimension 6 operators. 
}
PD operators of dimension 9 and higher, on the other hand, can involve \emph{three} leptons and thus have a richer flavor structure.
As a result they can conserve symmetries that are broken by lower-dimensional operators, leading to a dominance of the corresponding modes involving three leptons in the final states. For instance the $p\to e^+ e^+ \mu^-$ mode we will discuss at length below conserves $B-L$, $L_\tau$, and $L_e+2 L_\mu$. This decay can clearly not be brought back to other $B-L$-conserving modes such as $p\to \ell^+ \pi^0$ by closing SM loops, as this would require lepton-flavor-violating couplings. 

In the following we will determine the three-lepton dimension 9 and 10 operators arising in this way from a lepton flavor symmetry and identify the corresponding dominant nucleon decay channels. We will also present an example of a UV-complete leptoquark (LQ) model leading to the titular decays, which can furthermore accommodate neutrino masses and leptogenesis and also addresses the recent anomalies in $b\to s\mu\mu$ transitions.

\section{Dimension 9 operators}

The lowest operators with three quarks and three leptons have $d=9$. Those with $\Delta B=\Delta L/3$ do not lead to nucleon decay since they contain charm or top quarks~\cite{Weinberg:1980bf};
this leaves operators with $\Delta B=-\Delta L$, which, using Fierz-like identities, can be written in terms of scalar bilinears only. We find
\begin{align}
{\cal O}_{1}^9\ &= (QQ)_{1} (\bar{L}\bar{L})_{1} (\ell d)\,, &
{\cal O}_{2}^9\ &= (QQ)_{1} (\bar{L}\ell) (\bar{L}d)\,, \nonumber\\
{\cal O}_{3}^9\ &= (QL)_{1} (\bar{L}d) (\bar{L}d) \,, &
{\cal O}_{4}^9\ &= (\bar{\ell}Q) (\bar{L}d) (\ell d)\,,\nonumber\\
{\cal O}_{5}^9\ &= (\bar{L}\bar{L}) (u d) (\ell d)\,, &
{\cal O}_{6}^9\ &= (\bar{L} u) (\bar{L} d) (\ell d)\,,\nonumber\\
{\cal O}_{7}^9\ &= (\bar{L} d) (\bar{L} \ell) (u d)\,, &
{\cal O}_{8}^9\ &= (\bar{L} d) (\bar{L} d) (\ell u)\,,\label{eq:dim9}\nonumber\\
{\cal O}_{9}^9\ &= (QL)_{3} ((\bar{L}d) (\bar{L}d))_{3} \,, &
{\cal O}_{10}^9\ &= (QL)_{1} (\bar{L}\bar{L})_{1} (d d)\,,\nonumber\\
{\cal O}_{11}^9\ &= (QL)_{3} (\bar{L}\bar{L})_{3} (d d)\,, &
{\cal O}_{12}^9\ &= (\bar{\ell}Q) (\bar{L}\ell) (d d)\,, \nonumber\\
{\cal O}_{13}^9\ &= (\bar{L}\bar{L}) (u\ell) (d d)\,, &
{\cal O}_{14}^9\ &= (\bar{L}u) (\bar{L}\ell) (d d)\,, \nonumber\\
{\cal O}_{15}^9\ &= (\bar{\ell}L) (\bar{L}d) (d d)\,,&
{\cal O}_{16}^9\ &= (\bar{\ell}\bar{\ell}) (\ell d) (d d)\,.
\end{align}
Here, $Q$ ($L$) denotes the left-handed quark (lepton) doublet and $u$, $d$, and $\ell$ the right-handed quarks and lepton fields. We omitted all generation indices and $\epsilon_{ijk}$~color contractions, but indicated in subscripts the size of the non-trivial $SU(2)_L$ multiplet the fermion bilinear forms. These operators give rise to the dominant nucleon decays of Tab.~\ref{Tab:dim9}; there are no three-body PD modes, but ${\cal O}_{1}^9$--${\cal O}_{9}^9$ give $n\to \ell^+_\alpha \ell^-_\beta \nu_\gamma$, on which there are limits from IMB~\cite{McGrew:1999nd}. The other operators require an $s$ quark to survive the color anti-symmetrization, which then yield four-body decay modes involving kaons to be dominant, including the fully-visible $n\to K^+ \ell^+_\alpha \ell^-_\beta \ell^-_\gamma$ and partly visible $N\to K \ell^+_\alpha \ell^-_\beta \nu_\gamma$  channels.

\begin{table}[t]
	\begin{tabular}{lcr}
		channel & limit/$\unit[10^{30}]{yrs}$ & \hspace{1ex} operators \\
		\hline
		$n\to \ell^+_\alpha \ell^-_\beta \nu_\gamma$ &  $79$--$257$~\cite{McGrew:1999nd} & ${\cal O}_{1}^9$--${\cal O}_{9}^9$\\
		$N\to K \ell^+_\alpha \ell^-_\beta \nu_\gamma $ &  -- & ${\cal O}_{8}^9$--${\cal O}_{14}^9$\\
		$n\to K^+ \ell^+_\alpha \ell^-_\beta \ell^-_\gamma$ &  -- & ${\cal O}_{15}^9$--${\cal O}_{16}^9$\\
		\hline
	\end{tabular}
	\caption{Nucleon decay channels via the $d=9$ operators of Eq.~\eqref{eq:dim9}. Here, $N = (p\ n)^T$ and $K=(K^+\ K^0)^T$.
	}
	\label{Tab:dim9}
\end{table}

In order for these operators/channels to dominate over the  $d=7$, $\Delta B=-\Delta L$ channels~\cite{Weinberg:1980bf}, 
they need to carry lepton \emph{flavor} numbers that the lower ones cannot have. We find the corresponding list of dominant decays to be
\begin{align}
	n &\to e^+ \mu^- \nu_{\mu,\tau}\,, & n &\to \mu^+ e^- \nu_{e,\tau} \,,\nonumber\\
	N &\to K e^+ \mu^- \nu_{\mu,\tau}\,, & N &\to K \mu^+ e^- \nu_{e,\tau} \,,\label{eq:dim9decays}\\
	n &\to K^+ e^+ \mu^- \mu^-\,, & n &\to K^+ \mu^+ e^- e^- \,.\nonumber
\end{align}
One can readily identify the conserved symmetries for each decay.
Note that water Cherenkov detectors such as SK basically cannot determine the electric charge of the lepton, nor observe the outgoing neutrino, making it impossible to distinguish some of these channels.

Several modes of Tab.~\ref{Tab:dim9} were already discussed to some degree in the literature because they arise in $SU(4)_C$ unification models~\cite{Pati:1983jk} and in the $R$-parity violating MSSM~\cite{Carlson:1995ji,Bhattacharyya:1998dt,Bhattacharyya:1998bx}. A recent discussion of the latter case can be found in Ref.~\cite{Faroughy:2014tfa}, where it is claimed that the kaon modes typically dominate.
The corresponding lifetimes for massless leptons are~\cite{Faroughy:2014tfa}
\begin{align}
\hspace{-1ex}\Gamma (n\to \ell^+_\alpha \ell^-_\beta \nu_\gamma) &\sim \frac{\beta_\text{h}^2 m_n^5}{6144 \pi^3 \Lambda^{10}} \simeq \frac{\left(\unit[320]{TeV}/\Lambda\right)^{10}}{\unit[3\times 10^{32}]{yrs}} \,,\\
\hspace{-1ex}\Gamma (N\to K \ell^+_\alpha \ell^-_\beta \nu_\gamma ) &\sim \frac{\left(\unit[100]{TeV}/\Lambda\right)^{10}}{\unit[3\times 10^{32}]{yrs}} \,,
\label{eq:dim9_decay_rate}
\end{align}
with the hadronic matrix element $\beta_\text{h} \simeq \unit[0.014]{GeV^3}$~\cite{Aoki:2017puj} and ignoring order-one prefactors that depend on the actual operator ${\cal O}_{j}^{9}/\Lambda^5$ and lepton masses. 
Direct searches for these decays are either non-existent or rather old, thus we strongly encourage SK to search for the modes of Tab.~\ref{Tab:dim9}, in particular the flavor channels of Eq.~\eqref{eq:dim9decays}.

\section{Dimension 10 operators}

\begin{table}[t]
	\begin{tabular}{lrr}
		channel & $(\Delta L_e,\Delta L_\mu)$ & limit/years\\
		\hline
		$p\to e^+ e^+ e^-$ & $(1,0)$ & $793\times 10^{30}$\\
		$p\to e^+ \mu^+ \mu^-$ & $(1,0)$  & $359\times 10^{30}$\\
		$p\to \mu^+ e^+ e^-$ & $(0,1)$  & $529\times 10^{30}$\\
		$p\to \mu^+ \mu^+ \mu^-$ & $(0,1)$  & $675\times 10^{30}$\\
		$p\to \mu^+ \mu^+ e^-$ & $(-1,2)$  &  $359\times 10^{30}$  \\
		$p\to e^+ e^+ \mu^-$ & $(2,-1)$  &  $529\times 10^{30}$ \\
		\hline
	\end{tabular}
	\caption{$90\%$~C.L.~limits on PD branching ratios into three charged leptons~\cite{McGrew:1999nd}. The middle column shows the lepton flavor quantum numbers violated in the decay.
	}
	\label{Tab:limits}
\end{table}

There are two classes of $d=10$ operators with three leptons: 1) $\Delta B=\Delta L$, which can give rise to the six PD channels $p\to \ell^+_\alpha \ell^+_\beta\ell^-_\gamma$ (Tab.~\ref{Tab:limits}); 2) $\Delta B=-\Delta L/3$ which lead to four-body decays such as $n\to \nu \nu \ell \pi^+$~\cite{Weinberg:1980bf}.  The former class is particularly spectacular because it involves only three particles in the final state, all of which are charged leptons. The sensitivity of neutrino detectors to such a final state is expected to be as good or even better than for the usual two-body decays. This was in particular the case 20 years ago~\cite{McGrew:1999nd}, the last time these channels were searched for.
Therefore we strongly encourage experiments such as SK to perform dedicated searches for these channels.

We want to especially emphasize this for the 2 channels where both anti-leptons have the same flavor, $p\to e^+e^+\mu^-$ and $p\to \mu^+ \mu^+ e^-$, because  they can be singled out by a symmetry, $L_e+2 L_\mu+x L_\tau$ and $2 L_e+ L_\mu+x L_\tau$, respectively (with arbitrary value of $x$). Some $d=9$ operators/decays (Eq.~\eqref{eq:dim9decays}) also conserve one of these flavor symmetries, but 
they break $B-L$ and conserve $B+L$, opposite to the $d=10$ operators.
Thus, depending on the particle content and/or symmetries of the UV physics at the origin of these operators, it is perfectly possible that only the $d=10$ operators would be generated, see the explicit example of UV model below.
Note that the 4 PD channels which involve 2 different anti-leptons (Tab.~\ref{Tab:limits}) cannot be singled out from the two-body decays where the (flavor singlet) $e^+e^-$ or $\mu^+\mu^-$ pair is replaced by a (flavor singlet) $\pi^0$.  

Considering $d=10$ operators without a covariant derivative, the operators relevant for the channels of Tab.~\ref{Tab:limits} involve a SM scalar doublet field $H$. We find:
\begin{align}
{\cal O}_{1,2}^{10}\ &=\ (QQ)_{1,1}\,(QL)_{1,3} \,(\bar{L}\ell\bar{H})_{1,3}\,,\nonumber\\
{\cal O}_{3,4}^{10}\ &=\ (QQ)_{1,1}\,(QL)_{1,3} \,(\bar{\ell}L{H})_{1,3}\,,\nonumber\\
{\cal O}_{5}^{10}\ &=\ (QQ)_1\,(LL)_3 \,(\bar{\ell} Q {H})_3\,,\nonumber\\
{\cal O}_{6}^{10}\ &=\ (QQ)_1\,(\ell\ell)_1 \,(\bar{\ell} Q \bar{H})_1\,,\nonumber\\
{\cal O}_{7}^{10}\ &=\ (QQ)_1\,(LL)_3 \,(\bar{L} u {H})_3\,,\nonumber\\
{\cal O}_{8}^{10}\ &=\ (QQ)_1\,(\ell\ell)_1 \,(\bar{L} u \bar{H})_1\,,\nonumber\\
{\cal O}_{9}^{10}\ &=\ (QQ)_1\,(u\ell)_1 \,(\bar{L} \ell \bar{H})_1\,,\nonumber\\
{\cal O}_{10}^{10}\ &=\ (QQ)_1\,(u \ell)_1 \,(\bar{\ell} L {H})_1\,,\nonumber\\
{\cal O}_{11,12}^{10}\ &=\ (QL)_{1,3}\,(QL)_{3,3} \,(\bar{\ell} Q {H})_{3,3}\,,\nonumber\\
{\cal O}_{13,14}^{10}\ &=\ (QL)_{1,3}\,(QL)_{3,3} \,(\bar{L} u {H})_{3,3}\,,\nonumber\\
{\cal O}_{15,16}^{10}\ &=\ (QL)_{1,3}\,(u\ell)_{1,1} \,(\bar{\ell} Q {H})_{1,3}\,,\nonumber\\
{\cal O}_{17,18}^{10}\ &=\ (QL)_{1,3}\,(d\ell)_{1,1} \,(\bar{\ell} Q \bar{H})_{1,3}\,,\\
{\cal O}_{19}^{10}\ &=\ (QL)_{3}\,(u\ell)_{1} \,(\bar{L} u {H})_{3}\,,\nonumber\\
{\cal O}_{20,21}^{10}\ &=\ (QL)_{1,3}\,(d\ell)_{1,1} \,(\bar{L} u \bar{H})_{1,3}\,,\nonumber\\
{\cal O}_{22,23}^{10}\ &=\ (QL)_{1,3}\,(u\ell)_{1,1} \,(\bar{L} d \bar{H})_{1,3}\,,\nonumber\\
{\cal O}_{24,25}^{10}\ &=\ (QL)_{1,3}\,(ud)_{1,1} \,(\bar{L} \ell \bar{H})_{1,3}\,,\nonumber\\
{\cal O}_{26,27}^{10}\ &=\ (QL)_{1,3}\,(ud)_{1,1} \,(\bar{\ell} L {H})_{1,3}\,,\nonumber\\
{\cal O}_{28}^{10}\ &=\ (LL)_3\,(ud)_{1} \,(\bar{\ell} Q {H})_{3}\,,\nonumber\\
{\cal O}_{29}^{10}\ &=\ (ud)_1\,(\ell\ell)_{1} \,(\bar{\ell} Q \bar{H})_{1}\,,\nonumber\\
{\cal O}_{30}^{10}\ &=\ (u\ell)_1\,(d\ell)_{1} \,(\bar{\ell} Q \bar{H})_{1}\,,\nonumber\\
{\cal O}_{31}^{10}\ &=\ (LL)_3\,(ud)_{1} \,(\bar{L} u {H})_{3}\,,\nonumber\\
{\cal O}_{32}^{10}\ &=\ (ud)_1\,(u\ell)_{1} \,(\bar{L} \ell \bar{H})_{1}\,,\nonumber\\
{\cal O}_{33}^{10}\ &=\ (ud)_1\,(\ell\ell)_{1} \,(\bar{L} u \bar{H})_{1}\,,\nonumber\\
{\cal O}_{34}^{10}\ &=\ (u\ell)_1\,(d\ell)_{1} \,(\bar{L} u \bar{H})_{1}\,,\nonumber\\
{\cal O}_{35}^{10}\ &=\ (ud)_1\,(u\ell)_{1} \,(\bar{\ell} L {H})_{1}\,,\nonumber\\
{\cal O}_{36,37}^{10}\ &=\ (QL)_{1,3}\,(QL)_{1,3} \,(\bar{\ell} Q {H})_{1,1}\,,\nonumber\\
{\cal O}_{38,39,40}^{10}\ &=\ (QL)_{1,1,3}\,(QL)_{1,3,3} \,(\bar{L} d \bar{H})_{1,3,1}\,,\nonumber\\
{\cal O}_{41}^{10}\ &=\ (u\ell)_{1}\,(u\ell)_{1} \,(\bar{l} Q H)_{1}\,,\nonumber\\
{\cal O}_{42}^{10}\ &=\ (u\ell)_{1}\,(u\ell)_{1} \,(\bar{L} d \bar{H})_{1}\,,\nonumber
\end{align}
where the last 7 operators are only relevant for the channels $p\to e^+\mu^+ e^-$ and $p\to e^+\mu^+ \mu^-$.

With the above operators ${\cal O}_{j}^{10}/\Lambda^6$ we can calculate the induced PD rate, which for massless leptons is simply~\cite{Faroughy:2014tfa}
\begin{align}
\Gamma (p\to \ell^+_\alpha \ell^+_\beta\ell^-_\gamma) \sim \frac{\langle H\rangle^2 \beta_\text{h}^2 m_p^5}{6144 \pi^3 \Lambda^{12}} \simeq \frac{\left(\unit[100]{TeV}/\Lambda\right)^{12}}{\unit[10^{33}]{yrs}} \,.
\label{eq:decay_rate}
\end{align}
Judging by the limits on other three-body PDs~\cite{Chen:2014ifa,Takhistov:2014pfw}, a lifetime of this order is in reach of SK, thus probing  scales $\sim\unit[100]{TeV}$. The mediator masses in a UV-complete model can be even lower than this scale, since $\Lambda$ is also suppressed by couplings.
$SU(2)$-related PDs into less-visible modes such as $p\to \ell^+ \nu_{\ell'} \overline{\nu}_{\ell''}$ have been discussed in Ref.~\cite{ODonnell:1993kdg} but are of no interest here.

To reiterate, the PD channel $p\to e^+ e^+\mu^-$ ($\mu^+ \mu^+e^-$) could be dominant over all commonly discussed modes, as it is described by the lowest-dimensional operator that conserves $B-L$, $L_\tau$, and $L_e+2L_\mu$ ($L_\mu+2L_e$).
An analogous symmetry argument can be used to forbid PD operators up to $d=12$, only allowing, for example, for the PD operator $uud eee \bar \mu\bar \mu/\Lambda^8$. This leads to a PD scale as low as $\Lambda\sim \unit[10]{TeV}$.

\section{UV completion}

\begin{figure}[t]
	\includegraphics[width=0.42\textwidth]{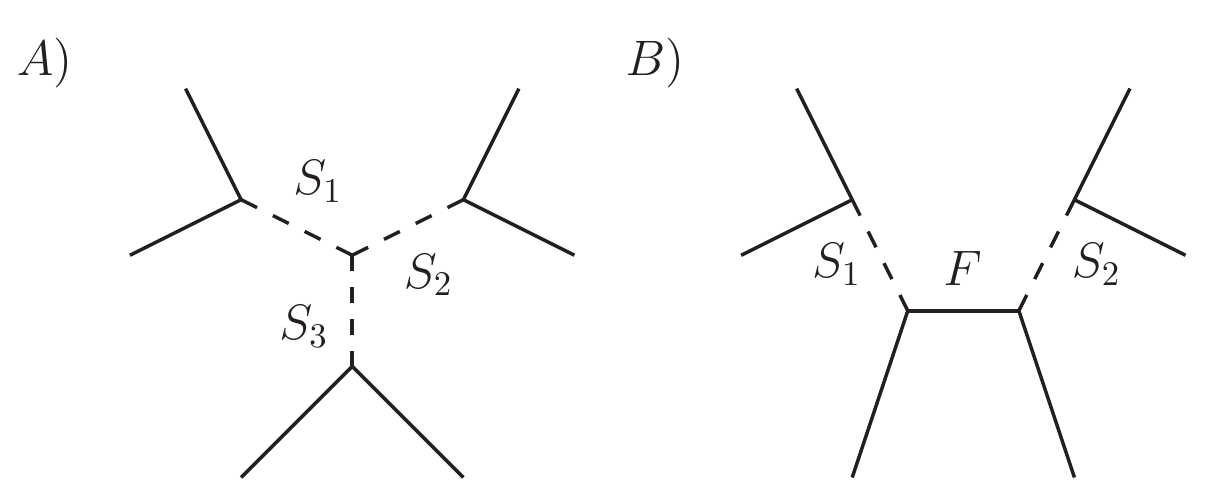}
	\caption{
		Topologies relevant for nucleon decay into three leptons. The external lines are labeled by three quarks and three leptons, which fixes the $SU(3)\times U(1)_\text{EM}$ charges of the internal scalars $S_j$ and fermion $F$.
	}
	\label{fig:proton_decay_topologies}
\end{figure}

Nucleon decay into three leptons via the $d=9$, $10$ operators discussed above can at tree-level proceed through the exchange of heavy particles along 2 different types of topologies, see Fig.~\ref{fig:proton_decay_topologies}. Topology $A$ involves new heavy scalars, whereas $B$ also involves a new heavy fermion. Emission of a kaon involves an extra spectator quark that does not change the discussion. (We omit an analogous discussion involving spin-1 mediators.)
For the $d=10$ operators there are various places in the diagram where the SM doublet $H$ can be inserted: on an external leg, on an internal propagator or on the trilinear scalar coupling in the diagram with topology $A$, making it a quartic coupling. We will not list explicitly all these possibilities, but instead give the possible quantum numbers of the heavy particles for all these possibilities.

First, the scalars along both topologies always couple to 2 SM fermions, and thus must have the corresponding quantum numbers. One finds that they are either $SU(2)_L$ singlet di-quarks (coupling to $\bar{Q}^c Q$, $\bar{u}^c d$, $\bar{d}^c d$), di-leptons (coupling to $\bar{\ell}^c \ell$, $\bar{\ell} L$, $\bar{L}^c L$) or LQs~\cite{Buchmuller:1986zs,Dorsner:2016wpm} (coupling to 
$\bar{\ell} d^c$, $\bar{\ell} u^c$, $\bar{L} Q^c$, $\bar{u} L$, $\bar{Q} \ell$, $\bar{d} L$), see also~\cite{KlapdorKleingrothaus:2002yq}. 
For the processes above involving a kaon, one of the $Q$ or $d$ quark field is intended to be of second generation. 
The present LHC lower bounds on the masses of these particles typically lie within $1$--$\unit[1.5]{TeV}$ for LQs and around $6$--$\unit[7]{TeV}$ for di-quarks~\cite{CMS:2017xrr}. 

As for the heavy fermion appearing in the diagram with topology $B$, it can be an $SU(3)_C$ singlet with electric charge $0$ or $1$ or a triplet with electric charge possibly equal to any multiple of 1/3 between $-7/3$ and $7/3$ except for $0$, $\pm1$, and $\pm 2$. Under $SU(2)_L$ all these particles can be singlet, doublet or triplet, depending in particular for the $d=10$ operators on where the Higgs doublet insertion is in the diagram. 
For more specific predictions we now turn to a UV complete example.

\subsection{Connection to \texorpdfstring{$b\to s$}{b to s} anomalies}

\begin{figure}[t]
	\includegraphics[width=0.3\textwidth]{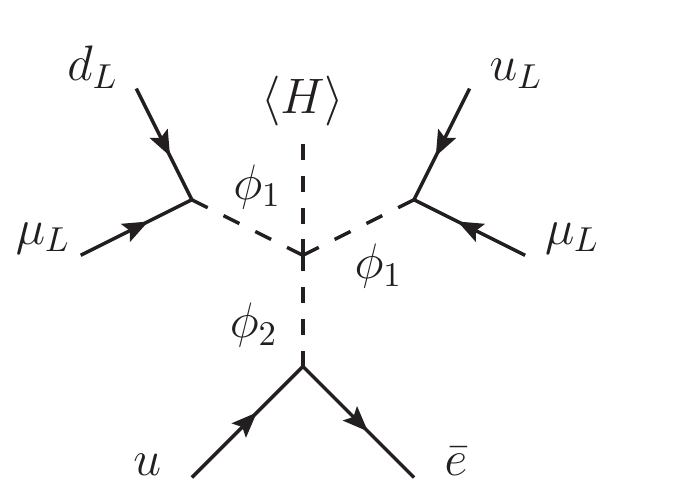}
	\caption{
		Diagram for  $p\to \mu^+ \mu^+ e^-$ with the LQs of Eq.~\eqref{eq:UV_lagrangian}, taking all fermions as incoming.
	}
	\label{fig:proton_S3}
\end{figure}

As a minimal model for $p\to \mu^+ \mu^+ e^-$ we take two LQs
\begin{align}
\phi_1 \sim (\vec{3},\vec{3},-2/3)\,, &&
\phi_2 \sim (\vec{3},\vec{2},7/3)\,,
\label{eq:UV_particles}
\end{align}
and assign them lepton flavors $L_\mu (\phi_1)=1=-L_e (\phi_2)$. Imposing a global (or even local) $U(1)_{L_\mu+2 L_e - 3 L_\tau}$~\cite{Araki:2012ip} restricts the relevant couplings in the Lagrangian to
\begin{align}
\hspace{-1ex} y_j \overline{L}_\mu \phi_1 Q_j^c + k_j \overline{Q}_j\phi_2 e  + f_j \overline{u}_j\phi_2 L_e  + \lambda \phi_1^2 \phi_2 H +\hc ,
\label{eq:UV_lagrangian}
\end{align}
$j$ being a quark-generation index. $B-L$ and $L_\tau$ are accidentally conserved (assigning $B(\phi_j)=1/3$).
Integrating out the heavy LQs yields the two PD operators
\begin{align}
\frac{\lambda y_1^2 k_1}{m_{\phi_1}^4 m_{\phi_2}^2} {\cal O}_{12,\mu\mu \bar{e}}^{10} \,,&&
\frac{\lambda y_1^2 f_1}{m_{\phi_1}^4 m_{\phi_2}^2}	{\cal O}_{14,\mu\mu \bar{e}}^{10} \,,
\end{align}
see Fig.~\ref{fig:proton_S3},
from which we can readily read off the suppression scale $\Lambda$ that gives the PD rate in Eq.~\eqref{eq:decay_rate}. Observable PD requires $m_{\phi_{1,2}}\simeq\unit[100]{TeV}$ for $\mathcal{O}(1)$ couplings.

Of course, integrating out the LQs not only gives $d=10$ operators, but also $d=6$ four-fermion operators such as $(\overline{L}_\mu Q^c_j)(Q_i L_\mu) y_j \overline{y}_i/m_{\phi_1}^2$, which conserve baryon number and lepton flavor on account of the $U(1)_{L_\mu+2 L_e - 3 L_\tau}$ symmetry. They \emph{do} however induce lepton-flavor \emph{non-universality}, which is an interesting signature in its own right. Limits from rare meson decays typically give limits on the operator effective scale $\Lambda$ of order $\unit{TeV}$ up to almost $\unit[100]{TeV}$, depending strongly on the quark-coupling structure~\cite{Buras:2014zga}. For couplings of order one and LQ masses around $\unit[100]{TeV}$, PD then easily dominates over low-energy constraints. This is even more true for smaller couplings, as the $d=6$ ($d=10$) operators are quadratic (linear) in the Yukawa couplings.

Focusing for example on the first-quark-generation couplings $\{y_1,k_1,f_1 \}$ relevant for PD, the only effects will be in pion decays~\cite{Bryman:2011zz,Campbell:2008um}, with $\phi_1$ ($\phi_2$) mediating decays into muons (electrons). The LQ contribution interfere with the SM in both cases~\cite{Dorsner:2016wpm}, which could be used to soften the bounds. Furthermore, it is also possible to keep $\Gamma (\pi^-\to e^- \overline{\nu}_e)/\Gamma (\pi^-\to \mu^- \overline{\nu}_\mu)$ SM-like by modifying both rates by the same amount. Without using any of these tricks we find the limits $m_{\phi_1}/y_1\gtrsim \unit[3]{TeV}$ and $m_{\phi_2}/\sqrt{k_1 f_1}\gtrsim \unit[200]{TeV}$, which easily allow for PD rates in reach of SK. The limit on the non-chiral LQ $\phi_2$ is particularly strong, but note that PD can proceed even if $k_1 f_1 =0$ as long as not both $k_1$ and $f_1 $ are zero.

While pion decays seem to satisfy lepton flavor universality, there are increasing hints for a violation in $B$-meson decays, specifically as a modification of $b\to s \mu^+\mu^-$~\cite{Aaij:2014ora,Aaij:2015oid}. The most recent addition here comes from LHCb as a smaller-than-SM value for $R(K^*) = \text{BR} (B^0\to K^{0*}\mu^+\mu^-)/\text{BR} (B^0\to K^{0*}e^+e^-)$~\cite{Aaij:2017vbb}. There is only one \emph{scalar LQ} representation that can explain all $b\to s$ data in addition to $R(K^{(*)})$ at tree level, which happens to be $\phi_1$ above~\cite{Hiller:2014yaa,Gripaios:2014tna,Varzielas:2015iva,Sumensari:2017ovu}; the resulting Wilson coefficient $C_9^\mu =-C_{10}^\mu$ improves the global fit by $4$--$5\sigma$ for  $m_{\phi_1}/\sqrt{y_2 y_3}\simeq \unit[30]{TeV}$~\cite{Altmannshofer:2017fio,Altmannshofer:2017yso,Alok:2017jgr,Alok:2017sui,Dorsner:2017ufx,Capdevila:2017bsm,DAmico:2017mtc}. 
Note that in this case one typically has to introduce a baryon symmetry to forbid the unwanted coupling $QQ\phi_1$ that would lead to fast PD~\cite{Dorsner:2016wpm}. In our scenario this is taken care of by the flavor symmetry, which furthermore ensures that $\phi_1$ only couples to muons, as required for the $b\to s$ data. Our symmetry is thus well-suited for the $b\to s$ anomalies independently of PD considerations.
With $y_2 y_3$ fixed to explain $b\to s$ data, new processes involving $b\to d$ and $s\to d$ transitions open up for $y_1\neq 0$, which need to be considered. A particularly constraining decay channel is $K^-\to \pi^- \nu_\mu \overline{\nu}_\mu$, which yields a limit $m_{\phi_1}/\sqrt{y_1 y_2}\simeq \unit[60]{TeV}$ that is easily compatible with observable PD.

\subsection{Neutrino masses and flavor symmetry breaking}

The UV model presented above shows explicitly that the $p\to \mu^+ \mu^+e^-$ channel can indeed be singled out and realized in a renormalizable model.
However, in order to allow for neutrino oscillations, the symmetry $U(1)_{L_\mu+2 L_e - 3 L_\tau}$ it involves must of course be broken, either softly or spontaneously. 
Let us introduce three right-handed neutrinos $N_{e,\mu,\tau}$ that carry the corresponding flavor charges; Dirac neutrino masses $m_D$ are then clearly allowed by the $U(1)$ symmetry, but mixing and Majorana masses for the $N_\alpha$ are still forbidden at this level. Introducing SM-singlet scalar fields $S_j$ with specific $U(1)_{L_\mu+2 L_e - 3 L_\tau}$ charges make it possible to write down Yukawa couplings $S_j \overline{N}_\alpha^c N_\beta$ that lead to a Majorana mass matrix $M_R$ which break the $U(1)_{L_\mu+2 L_e - 3 L_\tau}$ symmetry upon $S_j \to \langle S_j \rangle$. The structure in $M_R$ depends on the $S_j$ charges and could even lead to texture zeros~\cite{Araki:2012ip}, but the important point here is that it leads to neutrino oscillations, since in this case the $\propto m_D M_R^{-1} m_D$ seesaw mass matrix for the active neutrinos involve non-diagonal $M_R$.

Since the $U(1)_{L_\mu+2 L_e - 3 L_\tau}$ breaking occurs entirely in the SM-singlet sector, it does not have an impact on the above $p\to \mu^+ \mu^+e^-$ discussion; one can easily convince oneself that the $S_j$ vacuum expectation values will not be transfered to the $\phi_j$, so that the symmetry protection of $p\to \mu^+ \mu^+e^-$ is still in place. $U(1)_{L_\mu+2 L_e - 3 L_\tau}$-breaking processes such as $p\to \mu^+ \pi^0$ only arise with exchange of $N_j$, $S_j$ or $\nu_j$ \emph{on top of the diagram of Fig.~\ref{fig:proton_S3}}, which is heavily suppressed for large right-handed neutrino masses.

In this framework leptogenesis can proceed as usual, with $N_R$ decays at a high scale $M_R \sim\langle S_j \rangle$ providing a lepton asymmetry (both in total lepton number and our flavor $U(1)$) that is then transfered to baryons by sphalerons. The crucial observation here is that after the $N_R$ go out of equilibrium, our $U(1)_{L_\mu+2 L_e - 3 L_\tau}$ is conserved again, as well as $B-L$. This is sufficient to enable leptogenesis.

\section{Conclusion}
\label{sec:conclusion}

Proton decay is one of the most sensitive probes of physics beyond the SM. Given the stringent existing bounds, this typically forces new physics to conserve baryon number altogether. However, since PD unavoidably violates lepton flavor, it is possible that the dangerous (e.g.~two-body) channels would be forbidden on the basis of lepton flavor symmetries. These flavor symmetries must unavoidably be broken to allow for neutrino oscillations but this is practically irrelevant for proton decay. For example, the $d=9$ processes of Eq.~\eqref{eq:dim9decays} or the $d=10$ decays $p\to e^+ e^+ \mu^-$ and $p\to \mu^+ \mu^+ e^-$ processes could be the dominant PD modes. 
The last 2 channels in particular could be probed in SK for lifetimes up to few~$\unit[10^{34}]{yrs}$. Thus we encourage experimentalists to analyze their data for all these modes, which probe UV energy scales around \unit[100]{TeV}. These scales are low enough to potentially leave an impact in collider or meson decay observables. In fact, UV completions involving leptoquarks unavoidably induce lepton flavor non-universality and can nicely fit to recent hints for anomalies in $B\to K^{(*)} \mu^+\mu^-/B\to K^{(*)} e^+e^-$. 

\section*{Acknowledgements}

We thank Andreas Crivellin, Dario M\"uller, and Michele Frigerio for useful discussions.
This work is supported by the F.R.S.-FNRS, an ULB-ARC grant and the Belgian Federal Science Policy through the Interuniversity Attraction Pole P7/37.
JH is a postdoctoral researcher of the F.R.S.-FNRS.

\bibliographystyle{utcaps_mod}
\bibliography{BIB}

\end{document}